\documentstyle[prd,aps,preprint]{revtex}
\begin{document}
\draft

%
%  Uncomment following two lines and one below for 2 column format.
%@twocolumnfalse\endcsname

\preprint{Nisho-99/2} \title{A Possible Origin of Gamma Ray Bursts\\ 
and Axionic Boson Stars} 
\author{Aiichi Iwazaki}
\address{Department of Physics, Nishogakusha University, Shonan Ohi Chiba
  277,\ Japan.} \date{March 15, 1999} \maketitle
\begin{abstract}
We indicate a possible mechanism of generating gamma ray bursts.
They are generated by a collision between an axionic boson star and 
a neutron star. The axionic boson star can dissipates its whole energy
$10^{50}$ erg in the magnetized conducting medium of the neutron star. 
This dissipation arises only around envelope of the neutron star so that
a fire ball with small baryon contamination is expected.
We have evaluated roughly a rate of the collision per year and per galaxy
which is consistent with observations, under plausible assumptions. 
We also point out that cosmic rays with extremely high energy, $10^{21}$eV,
can be produced in the similar collisions with the neutron stars with 
strong magnetic fields $\sim 10^{14}$G.
\end{abstract}
%\pacs{73.61.-r,73.20.Dx,73.40.Hm,73.40.Gk}
\pacs{14.80.Mz, 98.80.Cq, 97.60.Jd, 98.70.Rz, 95.30.+d, 05.30.Jp, 
98.70.-f \\Axion, Gamma Ray Burst, Dark Matter, Boson Star 
\hspace*{3cm}}
\vskip2pc
%%%%%%%%%%%%%%%%%%%%%%%%%%%%%%%%%%%%%%%%%%%%%%%%
One of the most fascinating problems in astrophysics is the mechanism
leading to the gamma ray bursts ( GRB ). It is found that GRB can be 
understood well with the use of the fire ball models\cite{fb};
a relativistic expanding shell originating from the source interacts 
with the intergalactic matter and GRB is produced with the interactions.
However, we make clear the mechanism generating the fire ball.
Common understanding is that a merger of two compact stars, e.g. 
neutron stars or black holes, is the origin of the fire ball.
Especially a merger of two neutron stars is a plausible candidate.
But up to now, there is no satisfactory mechanisms generating the fire balls
with few contamination of baryons\cite{ba}.

In this letter we wish to propose a mechanism of generating GBR. 
The axionic boson stars ( ABS )\cite{real,iwaza}, 
which are ones of plausible candidates of the dark matters, collide with 
neutron stars and dissipate their energies 
under strong magnetic fields 
of the neutron stars. As a result, the whole energy of the axionic boson 
star is radiated very rapidly. Since the dissipation occurs 
in magnetospheres and around envelopes of the neutron star\cite{star}, 
the contamination problem of baryons can be cured.
Furthermore, rate of such collisions is estimated roughly as 
$10^{-6}\sim 10^{-7}$ per year in a galaxy under plausible
assumptions about number of neutron stars in the galaxy, e.t.c.. 
This rate is consistent with observations. A significant problem 
in this mechanism is that the maximal energy released is restricted 
as mass, $M_a$, of ABS itself; the mass is given by 
$\sim 10^{-5}M_{\odot}/m_5\sim 10^{49}/m_5$ erg 
where $m_5$ denotes $m_a/10^{-5}$eV with mass $m_a$ of the axion.
Since the mass of the axion is restricted\cite{text} observationally 
such as $m_a>10^{-6}$eV,
the maximal energy $E_m$ released in the collision must satisfy an upper bound 
such as $E_m<10^{50}$erg. This fact contradicts 
with a recent observation\cite{grb}
of GRB with the energy output up to $\sim 10^{54}$erg, 
which is estimated with assumption of no beaming. Since the beaming can 
occur in our model because of the presence of the strong magnetic field,
we expect that this energy problem may be cured.

First we explain briefly the axion and the axionic boson star.
The axion is the Goldstone boson associated with 
Peccei-Quinn symmetry\cite{PQ}, 
which was introduced to solve naturally the strong CP problem. 
The symmetry is broken at the energy scale of $f_{PQ}>10^{10}$GeV.
The resultant axion is described with a real scalar field, $a(x)$.
In the early Universe some of the axions
condense and form topological objects\cite{kim,text}, i.e. 
strings and domain walls, 
although they decay below the temperature of QCD phase transition.  
It have been shown numerically\cite{kolb} that although these
local objects decay,  
axion clumps are formed around the period of 
$1$ GeV owing to both the nonlinearity of an axion potential leading to 
an attractive force among the axions and 
the inhomogeneity of coherent axion oscillations\cite{text} 
on the scale beyond the horizon.    
The axion clumps contract gravitationally to 
axionic boson stars\cite{kolb2,real}
after separating out from the cosmological expansion.
They are solitons of coherent axions bounded gravitationally. 
They are described as stable solutions of the axion field coupled with 
gravity. 

It has been shown\cite{real} that such solutions exist and a critical mass 
of ABS is given by $\sim m_{pl}^2/m_a$; $m_{pl}$ is the Planck mass.
Quite similar results on the critical mass have been obtained for 
the boson stars\cite{re} of the complex scalar field; the critical mass 
implies the maximal mass
of the boson stars, with masses beyond which the stable solutions do not exit.
This is the same notion as the critical mass of the neutron stars.
Thus it is reasonable to suppose that as an order of magnitude,
a typical mass of ABS present in 
the Universe is given roughly by the critical mass,
$m_{pl}^2/m_a\sim 10^{-5}M_{\odot}/m_5\sim 10^{49}/m_5$ erg.
The solution can be approximated\cite{iwaza} in the explicit form such as

\begin{equation}
\label{a}
a=f_{PQ}a_0\sin(m_at)\exp(-r/R_a)\quad, 
\end{equation}
where $t$ ( $r$ ) is time ( radial ) coordinate and 
$f_{PQ}$ is the decay constant of the axion. 
The value of $f_{PQ}$ is constrained\cite{text} from cosmological 
and astrophysical considerations\cite{text,kim} such as 
$10^{10}$GeV $< f_{PQ} <$ $10^{13}$GeV.
$R_a$ represents a radius of 
ABS which has been obtained\cite{iwaza} numerically 
in terms of the mass $M_a$ of ABS,

\begin{equation}
\label{R}
R_a=6.4\,\frac{m_{pl}^2}{m_a^2M_a}\quad,
\end{equation} 
in the limit of infinitely large radius.
In the similar limit the amplitude $a_0$ of the axion field
is given by

\begin{equation}
\label{a_0}
a_0=1.73\times 10^6 \frac{(10\mbox{cm})^2}{R_a^2}\,
\frac{10^{-5}\mbox{eV}}{m_a}\quad.
\end{equation}
Numerical solutions\cite{real} obtained without taking the limit possess other 
oscillation modes, but their amplitudes are much smaller than this 
one eq(\ref{a_0}). Similarly, 
the almost same relation as one in eq(\ref{R}) between $R_a$ 
and $M_a$ can be obtained 
without taking the limit. Thus we use the formulae in rough evaluation 
of the energy released per unit time in the collision 
between ABS and the neutron star.
There are several physical parameters poorly known about 
the neutron stars so that the only evaluations of the order of 
the magnitude are meaningful.   

We note that ABSs oscillate with the frequency of 
$m_a/2\pi=2.4\times 10^{9}m_5$ Hz. 
and that the radius $R_a$
of ABS is given by $16\,m_5^{-2}M_{5}^{-1}$cm 
where $M_{5}$ denotes $M_a/10^{-5}M_{\odot}$.
% This value of the radius 
%is consistent with previous evaluations\cite{real} 
%without the approximation of 
%taking the limit of the infinitely large radius.

Up to now these solutions of ABSs have been found in the axion field 
equation only with approximating a cosine potential of the axion
with a mass term.
It seems that the treatment is inconsistent in the case of the large 
amplitude $a_0\sim 10^6/m_5$.
But we can see that the effect 
of the potential term, $\sim m_a^2f_{PQ}^2\cos(a/f_{PQ})$,
is negligible compared with other terms, 
e.g. $(\partial a)^2\sim R_a^{-2}f_{PQ}^2a_0^2\sim m_a^2f_{PQ}^2a_0^2$
in the case of the large amplitude $a_0\gg 1$ ( $R_a\sim m_a^{-1}$ in 
the case of our concerns ). Thus the solution in eq(\ref{a}) may be
used as approximate one even in the equation including the cosine potential
of the axion.

We now proceed to explain that ABS
generates an electric field
in an magnetic field of the neutron star and consequently dissipates 
its energy in a conducting medium. 
The point is that the axion couples\cite{kim} with the electromagnetic fields
in the following way,

\begin{equation}
   L_{a\gamma\gamma}=c\alpha a\vec{E}\cdot\vec{B}/f_{PQ}\pi
\label{EB}
\end{equation}
with $\alpha=1/137$, where 
$\vec{E}$ and $\vec{B}$ are electric and magnetic fields respectively. 
The value of $c$ depends on the axion models\cite{DFSZ,hadron};
typically it is the order of one.

It follows from this interaction that Gauss law is given by  

\begin{equation}
\label{Gauss}
\vec{\partial}\vec{E}=-c\alpha \vec{\partial}\cdot(a\vec{B})/f_{PQ}\pi
+\mbox{``matter''}
\end{equation}
where the last term ``matter'' denotes contributions from ordinary matters.
The first term in the right hand side 
represents a contribution from the axion.
Thus it turns out that
the axion
field has an electric charge density, 
$\rho_a=-c\alpha\vec{\partial}\cdot(a\vec{B})/f_{PQ}\pi$, 
under the magnetic field $\vec{B}$\cite{Si}.
Accordingly, the electric field, $E_a$ associated with this axion charge
is produced such that 
$\vec{E_a}=-c\alpha a\vec{B}/f_{PQ}\pi$.  
Note that this field is quite strong around the surface of neutron star
with a magnetic field $10^{12}$G;
$E_a\sim 10^{18}\,(B_{12}/m_5)\,\,\mbox{eV}\,\,\mbox{cm}^{-1}$ 
with $B_{12}=B/10^{12}$G.

Note that both of $\rho_a$ and $E_a$ oscillate with the frequency given by 
the mass of the axion in ABS, since 
the field $a$ itself oscillates. The typical spatial extension 
of the electric field 
is about $10$cm, while the frequency of the oscillation is $10^9$ Hz 
for ABS of our concern. Thus the electric field can not be screened 
by charged particles present around the neutron stars.

Obviously,
this field induces an oscillating electric current $J_m=\sigma E_a$
in magnetized conducting media with electric conductivity $\sigma$. 
In addition to the current $J_m$ carried 
by ordinary matters, e.g. electrons, there appears 
an electric current, $J_a$, associated with the oscillating charge $\rho_a$
owing to the current conservation\cite{Si} 
( $\partial_0\rho_a-\vec{\partial}\vec{J_a}=0$ ). 
This is given such that 
$\vec{J_a}=-c\alpha\partial_{t}a\vec{B}/f_{PQ}\pi$.  
This electric current is present even in nonconducting media like 
vacuum as far as ABS is exposed to the magnetic field.
On the other hand, current $J_m$ is present 
only in the magnetized conducting media.

Since $\partial_t a\sim m_aa$ in ABS, 
the ratio of $J_m/J_a$ is given by $\sigma/m_a$. Hence, 
$J_a$ is dominant in the media with $\sigma < 10^{12}/s$, while $J_m$
is dominant in the media with $\sigma > 10^{12}/s$; 
note that $10^{9}/s < m_a < 10^{12}/s$ 
corresponding\cite{text} to the above constraint on $f_{PQ}$. 
Electric conductivities in the neutron stars are 
large enough for 
$J_m$ to be dominant, while the magnetospheres of the 
neutron stars may have the small conductivities 
so that $J_a$ is dominant. 
The envelopes of the neutron stars 
have still much large conductivities so that $J_m$ is dominant.
Therefore, we expect that as ABS approaches the neutron star, 
both of the magnetic field and the conductivity become large and hence 
the rate of the dissipation of its energy increases; the rate is 
proportional to $\sigma E_a^2$. 
We will see soon later that ABS dissipates quite rapidly its 
whole energy inside of the neutron star 
because of the extremely high electric conductivity.

Let us evaluate amount of the energy dissipated in the magnetized 
conducting medium such as the magnetosphere and the inside of the neutron
star. Denoting the average electric conductivity of the media 
by $\sigma$ and assuming 
the Ohm law, we find that
the axion star dissipates an energy $W$ per unit time, 

\begin{equation}
W=\int_{ABS}{\sigma E_a^2}d^3x=
4c^2\times 10^{54}\mbox{erg/s}\,\frac{\sigma}{10^{26}/s}\,
\frac{M}{10^{-4}M_{\odot}}\,\frac{B^2}{(10^{8}G)^2}
\end{equation} 
where the integration has been performed over volume of ABS and 
we have used the explicit formula of $a_0$ and $R_a$ in the above.
The value of the conductivity $\sigma$ is taken as a typical value of 
the inside of the neutron star, although its precise value depends on 
temperature, density, composition e.t.c. of the neutron star\cite{con}. 
When we consider magnetosphere with very low conductivity, $\sigma <10^9$/s,
we should replace conductivity, $\sigma$, with mass $m_a$. 

As we can see from the equation of $W$, the dissipation proceeds very rapidly 
in the neutron star even with relatively low magnetic field $B\sim 10^8$G.
Since the real neutron stars, even old ones, 
must possess stronger magnetic fields than $10^8$G,
ABS with mass $10^{-4}M_{\odot}\sim 10^{50}$erg
evapolates within a time less than $10^{-4}$ second in the core of the 
neutron star.
It means that the real 
dissipation arises only near the envelope of the neutron star; ABS can not 
enter the core of the neutron star since a typical velocity of ABS
in a halo is $10^{-3}\times \mbox{light velocity}$.
Therefore, the energy released with
the collision between ABS and the neutron star occurs around the surface of 
the neutron star. This implies that a fire ball, which 
would be produced with the collision, may 
involve only a small fraction of baryons. 
We expect that this is a mechanism of generating 
the fire ball with few baryon contamination and 
hence with large Lorentz factor. 
Although we have neglected the dissipation in the region of the magnetosphere,
even if the dissipation is so large for the whole energy of ABS to 
be dissipated in the region, 
the baryon contamination in the fire ball is smaller than one in the case of 
dissipation arising mainly near the envelope.

The amount of the energy released in this mechanism 
is given maximally by $10^{50}$ erg which is the mass,
$\sim m_{pl}^2/m_a$ of ABS 
with the choice of axion mass $m_a=10^{-6}$eV; possibility of 
the axion with the axion mass smaller 
than $m_a=10^{-6}$eV is inhibited cosmologically.  
Thus the maximal energy released is less than 
one observed in the gamma ray burst.
However, in our mechanism generating the clean fire ball we can expect beaming 
of gamma ray emissions because of the presence of the strong magnetic field 
of the neutron star. Therefore, the observations do not necessarily 
contradict with our mechanism.  

Here we wish to point out that since the electric field generated under
the magnetic field of the neutron star is quite strong,
$E_a\sim 10^{18}\,(B_{12}/m_5)$ eV $\mbox{cm}^{-1}$,
it is possible for the field to produce cosmic rays with extremely high 
energy such as $10^{21}$eV. Note that in a period $10^{-9}/m_5$ sec 
of the oscillation 
of the field, a charged particle accelerated by the field
can gain energy, 
$E_a\times 10^{-9}\mbox{sec}\times \mbox{light velocity}$.  
Thus, the energy gain such as $10^{21}$eV
is realized under the strong magnetic field
$\sim 10^{14}$G. Hence we speculate that 
the mechanism of generating GRB is the same as 
one generating extremely high energy cosmic rays, although 
the neutron star need to possess stronger magnetic field for 
the generation of such cosmic rays than one the ordinal neutron star does.

Finally we evaluate the rate of the collisions between the neutron stars
and ABSs. We assume that number of the neutron stars in a galaxy is 
$10^8\sim 10^9$ just as expected in our galaxy. 
We also assume that the halo of the galaxy
is composed mainly of ABS whose typical velocity, $v$ is supposed to be 
$3\times 10^7$ cm\cite{text}. 
Since local density of the halo in our galaxy is given by 
$0.5\times 10^{-24}$g $\mbox{cm}^{-3}$\cite{text},  
the rate, $R_c$ of the collision per year and per a galaxy 
is calculated as follows,

\begin{equation}
R_c=n_a\times N_{ns}\times Sv\times 1\,\mbox{year}
\simeq 10^{-7}m_5\sim 10^{-6}m_5\,\,
\mbox{per year}
\end{equation}
with $n_a=0.5\times 10^{-24}\mbox{g cm}^{-3}/M_a$ being number density of ABS
in the galaxy and $N_{ns}=10^8\sim 10^9$ being the number of the neutron stars.
Cross section, $S$ of the collision has been caluculated in the following.
Namely, we suppose that the collision occurs when these two objects approach
with each other such that 
the kinetic energy, $v^2M_a/2$ of ABS is equal to  
the potential energy $GM_aM_{\odot}/L_c$ of ABS around the neutron star
with mass $M_{\odot}$; $L_c\sim 10^{11}$ cm.
Thus $S=\pi L_c^2$.
This means that ABS is trapped to the neutron star when they approach 
within a distance less than $L_c$. 
But this estimation of $L_c$ is too naive. In a real situation of the 
trapping ABS, we need to take account of the dissipation of 
both the kinetic energy and the angular momentum of ABS 
in the region of the magnetosphere or in outer region 
with the magnetic field. In these regions ABS interacts with the magnetic 
field as mentioned above and looses its energy or angular momentum.   
Therefore, it turns out that the rate, $R_c$ 
depends on the several parameters poorly known,
e.g. electric conductivity of the surrounding of the neutron star. But
we may expect from the naive estimation 
that the rate of the collisions is not necessarily 
inconsistent with the observations of GRB.

There must be various ways of ABS being trapped; 
ABS moves around the neutron stars
several times before colliding the neutron star 
or it collides directly with the neutron stars. Thus we expect that
there are several types of pulse of GRB. Actually, various 
shapes of the pulses have been observed\cite{em}. 
These variations might originate from 
the ways of ABS being trapped.

In summary, we have proposed a possible origin of generating GRB; 
the collision between ABS and the neutron star. 
In the collision ABS evapolates
rapidly near the surface of the neutron star 
and maximally the energy $10^{50}$ erg can be  
released. We can expect quite a few contamination of baryons in 
a fire ball produced in this mechanism.
We have also evaluated the rate of the collisions in a 
typical galaxy and have found that it is roughly 
$10^{-7}m_5\sim 10^{-6}m_5$ per year,
although there are several ambiguities in the evaluation.   
Since there are various ways of the collisions depending on the collision 
parameters, the variations of the shapes of GRB observed are possibly 
caused by theses ways of ABS colliding with the neutron star.
Furthermore,
we have pointed out that cosmic rays with extremely high energy, $10^{21}$eV,
can be produced in the collisions between ABSs and the neutron stars with 
strong magnetic fields $\sim 10^{14}$G.

The author wish to express his thank for the hospitality in Tanashi KEK.

%%%%%%%%%%%%%%%%%%%%%%

\end{document}